\documentclass[psfig,showpacs,preprint,aps,prb]{revtex4}
\usepackage{graphicx}

\usepackage{graphicx,amsmath,amssymb,color}
\begin{document}
\def\litem{\par\noindent\hangindent=\parindent\ltextindent}
\def\ltextindent#1{\hbox to \hangindent{#1\hss}\ignorespaces}
\vskip 1.5truecm

\title{Dynamics of the Distorted Diamond Chain}

\author{H.-J. Mikeska and C. Luckmann}
\affiliation{Institut~f\"ur~Theoretische~Physik, Universit\"at~Hannover, 
30167~Hannover, Germany}

\date{\today} 

\begin{abstract}
We present results on the dynamics of the distorted diamond chain,
S=1/2 dimers alternating with single spins 1/2 and exchange couplings
$J_1$ and $J_3$ in between. The dynamics in the spin fluid (SF) and
tetramer-dimer (TD) phases is investigated numerically by exact
diagonalization for up to 24 spins. Representative excitation spectra
are presented, both for zero magnetic field and in the 1/3 plateau
phase and the relevant parameters are determined across the phase
diagram. The behaviour across the SF-TD phase transition line is
discussed for the specific heat and for excitation spectra. The
relevance of the distorted diamond chain model for the material
Cu$_3$(CO$_3$)$_2$(OH)$_2$ (azurite) is discussed with particular
emphasis on inelastic neutron scattering experiments, a recent
suggestion of one possibly ferromagnetic coupling constant is not
confirmed.
\end{abstract}


\maketitle

\section{Introduction}
\label{sec:intro}

\begin{figure}
\raisebox{15mm}{\includegraphics[width=80mm]{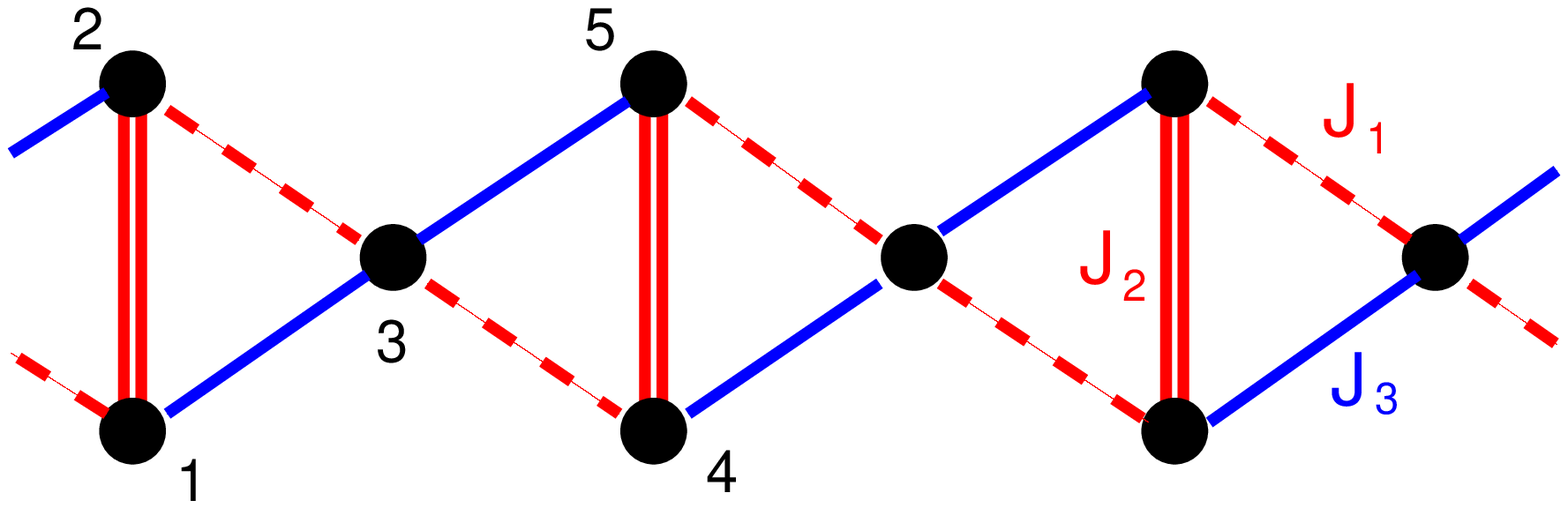}}
\hspace*{10mm}
\includegraphics[width=50mm]{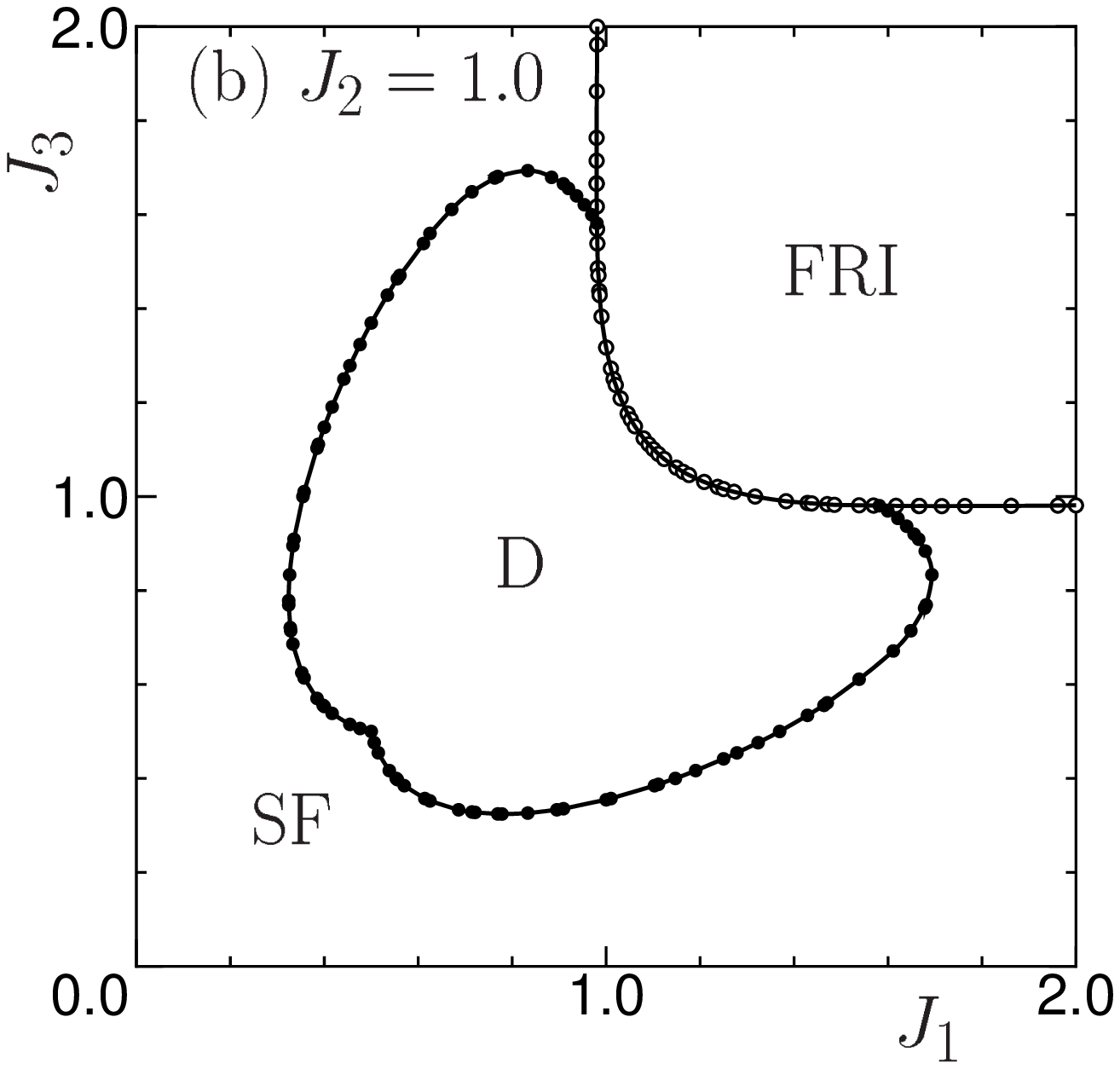}
\caption{(Color online) Structure (a) and phase diagram (b) of the distorted
diamond chain model ((b) from Ref.\onlinecite{TonO2000})}
\label{fig:structure+phases}
\end{figure}

The distorted diamond chain (DDC) is a one-dimensional (1D) quantum spin model
with structure as shown in Fig.~\ref{fig:structure+phases}(a) and
hamiltonian  
\begin{equation}
\label{eq:hamiltonian}
H = \sum_{n=1}^{N/3} \Big\{ J_2 \ \mathbf{S}_{3n+1} \mathbf{S}_{3n+2}
    + J_1 \ (\mathbf{S}_{3n} \mathbf{S}_{3n+1} 
                   + \mathbf{S}_{3n+2} \mathbf{S}_{3n+3}) 
    + J_3 \ (\mathbf{S}_{3n} \mathbf{S}_{3n+2} 
                   + \mathbf{S}_{3n+1} \mathbf{S}_{3n+3}) \Big\}. 
\end{equation}
This model with spins 1/2 and all couplings antiferromagnetic may be strongly
frustrated owing to the triangular building blocks and has receiced increasing
interest in the last decade for a number of reasons \cite{TakKS96}: It has a
rich quantum phase diagram as shown in Fig.~\ref{fig:structure+phases}(b)
(taken from Ref.~\onlinecite{TonO2000}). Here and in the following we choose a
representation with $J_2 = 1$ as energy unit and $J_1, J_3$ as variables. This
representation emphasizes the symmetry of the model under exchange of $J_1$
and $J_3$. Three quantum phases have been discussed for the ground state of
the model in zero magnetic field: For $J_1, J_3 \ll 1$ the ground state
develops from the state with dimers in their singlet state on $J_2$ bonds and
nearly free spins between these dimers. The low energy sector is governed by
an effective antiferromagnetic Heisenberg chain with $N/3$ sites resulting
from the residual coupling between these spins and denoted as $J_{\rm eff}$ in
the following. This leads to the formation of a spin fluid (SF) phase with
additional high energy excitations. For intermediate $J_1, J_3$ the ground
state dimerizes, forming a twofold degenerate sequence of alternating
tetramers and dimers (TD phase). Finally, for both $J_1, J_3$, sufficiently
large, the ground state is ferrimagnetic with e.g. a $\uparrow \uparrow
\downarrow$ structure of the unit cell of three spins (which satisfies $J_1$
and $J_3$ bonds and frustrates $J_2$). These three phases can be clearly
identified already in the symmetric model with $J_1=J_3$ in the regimes
$J_1=J_3 \le 1/2$ (SF phase), $1/2 \le J_1=J_3 \le J_m$ (TD phase) and $J_m
\le J_1=J_3$ (ferrimagnetic phase) with $J_m \approx 1.10$ \cite{TakKS96}.

The generalization to the distorted diamond chain $J_1 \ne J_3$ leads to an
even richer behaviour including e.g.~the trimer Heisenberg antiferromagnetic
chain on the $J_3=0$ axis with the standard HAF for $J_1=1$ as limiting
case. The DDC model can be seen as generalization of the HAF with NN and NNN
interactions: The critical point of this model at $J_{\rm NNN}/J_{\rm NN}
\approx 0.2411$ which marks the transition from a spin fluid to the dimerized
phase is extended into the line between the SF and the TD phases in the DDC
model and the Majumdar Ghosh point at $J_{\rm NNN}/J_{\rm NN} = 1/2$ is
extended into the line $1/2 \le J_1=J_3 \le J_m$ with simple and exactly known
dimerized ground states. In particular, the transition line between the SF and
TD phases is a line of Kosterlitz-Thouless phase transitions. The point $J_1 =
J_3 = 1/2$ is a point of particular high degeneracy. However, compared to the
simple HAF chain with nearest and next nearest exchange, the DDC model has the
dimer subsystem as an additional degree of freedom which dominates the high
energy regime, but also, in the appropriate parameter regime, interacts with
the low energy spin part. The ferrimagnetic part of the quantum phase diagram
is another consequence of the combined influence of these two degrees of
freedom.

An external magnetic field polarizes the quasi-free spins and at a critical
field $H_{c1}$ produces a magnetization plateau at 1/3 of the total saturation
magnetization. The plateau state corresponds to a fully saturated subsystem of
spins 1/2 and all dimers ($J_2$ bonds) in their singlet state. Further
increase of the moment requires breaking up at least one dimer with its large
energy scale $J_2 = 1$ and therefore a correspondingly higher field $H \ge
H_{c2}$ (end of the plateau). Finally complete saturation is obtained at the
field $H_{\rm sat}$, given by
\begin{equation} 
\frac{H_{\rm sat}}{J_2} = \frac{1}{2} + \frac{3}{4} (J_1 + J_3)
  + \frac{1}{2} \left\{ 2 (J_1-J_3)^2 + \frac{1}{4} (J_1+J_3)^2 +1 - (J_1 +
  J_3)\right\}^{\frac{1}{2}}.
\label{saturation}
\end{equation}

The critical field $H_{\rm c1}$ (beginning of the 1/3 plateau) is determined
by the level crossing between the saturated state of the effective HAF and its
'ferromagnon' excitation with one unit of magnetization less. If the mapping
to the effective HAF applies, this gives the relation
\begin{equation} 
H_{\rm c1} = 2 J_{\rm eff}. 
\label{criticalfield}
\end{equation} 

Apart from the theoretical interest in investigating a model which allows to
follow the variation between different quantum phases, the DDC model is of
interest since it seems to describe reasonably well the compound azurite,
Cu$_3$(CO$_3$)$_2$(OH)$_2$. Azurite has been investigated in detail by static
measurements (magnetization, susceptilibity, specific heat \cite{Kik05}) as
well as by high field ESR \cite{Oht03} and the existence of the 1/3
magnetization plateau has been clearly established.  From these experiments,
this compound appears to be in the SF phase close to the phase transition to
the TD phase. Recently, however, the possibility of one of the couplings $J_1,
J_3$ being ferromagnetic has been suggested from susceptibility and specific
heat data \cite{Hon07}. Beyond the static properties investigated so far, the
dynamics of the DDC and the material azurite in particular remain as a
challenge to be understood both experimentally and theoretically: the
characteristic feature of the model, namely the presence of two degrees of
freedom with different energy scales and their mutual influence will show up
most clearly in the energy spectra of the model. These are best investigated
by inelastic neutron scattering (INS) experiments as clearly seen in recent
work \cite{RulST07}. For a more complete description both of the DDC in the
full phase diagram and of the results of INS experiments on azurite we present
in the following results on the dynamics in the SF phase (section
\ref{sec:sf-dynamics}) and in the TD phase (section \ref{sec:td-dynamics}).

A perturbative approach can be applied to obtain results in the regime $J_1,
J_3 \ll 1$ (see Ref.~\onlinecite{HonL01}) as well as close to some special
points in the phase diagram.  Generally, however, for a quantitative
description numerical calculations are required. We will present in the
following the results of exact diagonalization, using both the Lanczos
algorithm for systems with 24 spins and complete diagonalization (all
eigenvalues) for 12 and 18 spins. The latter are necessary since the Lanczos
algorithm gives only a limited number of the lowest energy levels in the
subspace considered (in our case: $S^z_{\rm tot}$ and wave vector $k$) which
is not sufficient to cover the excited dimer subspace with its higher
energies. Since the elementary cell has 3 spins, our system sizes are
restricted to 4, 6 and 8 elementary cells. It turns out, however, that this
for many aspects is sufficient to obtain reliable results for the infinite
system when a finite size analysis is carried through.

\section{Dynamics in the Spin Fluid Phase}    
\label{sec:sf-dynamics}

It is helpful to start the discussion from two well known limiting cases:\\
(i) For $J_1=J_3=0$ the system reduces to $N/3$ independent dimers and $N/3$
free spins. In the ground state all dimers are in their singlet state and a
$2^{N/3}$~fold degeneracy due to the free spins results. A magnetic field
immediately saturates the free spin system leading to a magnetization of
$M_{\rm sat}/3$ which remains constant until at a field $H_{c2} = J_2$ the
dimers change to their triplet states saturating the system. This behavior to
a certain extent remains valid on the symmetry line $J_1 = J_3$ where the
total spins on all $J_2-$bonds are independently conserved: the ground state
as well as the 1/3 magnetization plateau in low field remain unchanged whereas
the transition to full saturation is determined by the effective interaction
which develops between two neighboring dimers in their triplet state and
finally leads to an effective $S=1$ chain. Qualitatively, for a large range
of parameters the distorted diamond chain can be divided into two subsystems
with clearly different energy scales, a low energy part of $N/3$ spins 1/2 and
a high energy part of $N/3$ dimers.

For small deviations from the independent free spin limit, $J_1 \ne J_3$, the
spin 1/2 subsystem develops some coupling by polarizing intermediate dimers
and the $2^{N/3}$~fold degeneracy is lifted in favor of an effective
Heisenberg chain with exchange $J_{\rm eff}$. In this regime, excitations of
the DDC remain well separated: they are in the low energy regime with energy
scale $J_{\rm eff}$ forming the spinon continuum of the HAF with $N/3$ spins
in the Brillouin zone of the full DDC (lattice constant $a$, reciprocal
lattice vector $\tau = 2\pi/a$) or in the high energy regime with energy scale
$J_2 = 1$ corresponding to the excitation of a dimer to its triplet state and
developping into a dispersive band with width $\Delta_{\rm dimer}$ due to the
coupling to the low energy spin subsystem. We will not consider in the
following states with more than one excited dimer.

(ii) For $J_1 = 1, J_3 = 0$ the system reduces to the Heisenberg
antiferromagnet with $N$ spins, forming a spinon continuum in the Brillouin
cell with reciprocal lattice vector $3\tau$, energy scale 1 and no additional
high energy excitations. For $0 < J_1 <1$ we have a trimerized Heisenberg
chain and the spectrum is obtained by folding back the spinon continuum to the
smaller Brillouin zone corresponding to lattice constant $a$. This results in
three excitation branches (actually continua) which fill the energy range up
to $\epsilon = \pi$ with small (for $J_1$ slightly less then 1) gaps between
them and an alternating sequence of minimum, maximum and minimum at wavevector
$k=\pi$ (in the following we will use exclusively the Brillouin zone with
reciprocal lattice vector $\tau$, corresponding to the full DDC). With
increasing $1-J_1$, these trimer bands develop increasingly larger gaps,
finally the continuum of the effective HAF emerges from the lowest band and
the two upper bands conspire to give the dimer excitations decorated by
continua of low energy spinon excitations.

Using this frame the lowest excitations of interest in the following are easily
described:\\ 
(i) the spinon continuum of the effective chain,\\
(ii) the band with one excited dimer above the spinon continuum,\\
(iii) the 'inverted ferromagnon' i.e.~the saturated effective HAF with one spin
deviation ($S_{\rm tot}^z= \frac{1}{2} N/3 - 1$), and\\
(iv) the band with one excited dimer above the saturated effective HAF (one
dimer in its triplet state ($S_{\rm tot}^z = \frac{1}{2} N/3 + 1$).\\
The dispersion of excitations (ii)-(iv) is determined by hopping processes
(spin deviations resp. dimer triplets moving to neighboring sites due to the
residual interactions). To lowest order these processes result in a cosine
dispersion and we introduce as notation for (ii)
\begin{equation}
\epsilon_{\rm dimer}^{\rm (0)}(k) = 
1 + \delta_{\rm dimer}^{\rm (0)} + \frac{1}{2} \ \Delta_{\rm dimer}^{\rm
       (0)} \cos k. 
\label{eq:spinonexciteddimer}
\end{equation}
More precisely, this excitation is not a single band but a continuum due to
the spinon continuum of initial states; however, we will only be able to
discuss the lower edge of this excitation and therefore simplify the notation
using Eq.(\ref{eq:spinonexciteddimer}). Excitations (iii) and (iv) are the
relevant excitations above the 1/3 plateau, we therefore use a notation giving
their energies in finite magnetic fields relative to the plateau ground state
with $S_{\rm tot}^z = \frac{1}{2} N/3$:
\begin{eqnarray}
\label{eq:ferromagnon}
\epsilon_{\rm ferrom}(k) &=&  
        \frac{1}{2} \ \Delta_{\rm ferrom} (1 + \cos k) + H - H_{\rm c1} \\  
\label{eq:satexciteddimer}
\epsilon_{\rm dimer}^{\rm (sat)}(k) &=& 1 \ + \
 \frac{1}{2} \ (J_1 + J_3) \ + \ \delta_{\rm dimer}^{\rm (sat)} + 
 \frac{1}{2} \ \Delta_{\rm dimer}^{\rm (sat)} (1 + \cos k) - (H - H_{\rm c1}).
\end{eqnarray}
The quantities $\Delta$ and $\delta$ give the widths, resp.~the nontrivial
contributions to the minimum energy (at $k = \pi$) of the corresponding
bands. The cosine dispersion of course is only valid in lowest order and will
change to a more complicated expression for real systems.

In the model of an effective HAF for the low energy regime its exchange
constant $J_{\rm eff}$ determines the low energy spinon (i) and the inverted
ferromagnon spectrum: $\Delta_{\rm ferrom} = 2 J_{\rm eff}$. Combined with
$\epsilon_{\rm dimer}^{\rm (sat)}$ it is also sufficient to give the range of
the plateau phase and to characterize its dynamics: In the presence of a
finite field, spectra are identical to those without field except for the
shifts and splittings due to Zeeman energies. This establishes states with an
increasingly larger total spin $S_{\rm tot}$ (in their maximum $z-$component)
as ground states. The plateau begins at the field $H_{\rm c1}$ when the
$S_{\rm tot}= \frac{1}{2} N/3$ level (saturated state of the quasi-free spin
subsystem) is forced below the lowest $S_{\rm tot}= \frac{1}{2} N/3 - 1$ level
('ferromagnon' band at wavevector $\pi$) by the external field, leading to
$H_{c1} = 2 J_{\rm eff}$.  The lowest excitation for the plateau dynamics
close to the field $H_{c1}$ then is the ferromagnon of
Eq.(\ref{eq:ferromagnon}). When the field is increased across the plateau
regime, the $S_{\rm tot}= \frac{1}{2} N/3 + 1$ level (one excited dimer on top
of the saturated quasi-free spin subsystem) lowers its energy, crosses the
ferromagnon excitation and finally is responsible for the end of the plateau
at the upper plateau field $H_{c2}$ implying
\begin{equation}
H_{\rm c2} = H_{\rm c1} + 1 + \frac{1}{2} (J_1 + J_3) 
                   + \delta_{\rm dimer}^{\rm sat}.
\end{equation}

The parameters determining the spectra can be calculated in perturbation
theory in $J_1, J_3$ and to lowest order are determined by the level spectrum
of the general ($J_1 \ne J_3$) tetramer with 4 spins 0 \dots 3. This spectrum
includes the lowest order information about $J_{\rm eff}$ in the
singlet-triplet splitting of spins 0 and 3 and about $\Delta_{\rm dimer}^{\rm
(sat)}$ in the amplitude for the process $\vert \uparrow s \rangle \to \vert
\downarrow t_+ \rangle$ ($s$ and $t_+$ are noninteracting dimer states) which
determines the propagation of an excited dimer triplet. The results to lowest
order in $J_1, J_3$ are:
\begin{eqnarray}
2 J_{\rm eff} = \Delta_{ferrom} = 2 \Delta_{\rm dimer}^{\rm sat} 
   &=& (J_1 - J_3)^2 \nonumber \\
\delta_{\rm dimer}^{\rm sat}  &=& -(J_1 - J_3)^2 \nonumber \\
H_{c2} &=& 1 + \frac{1}{2} (J_1 + J_3). 
\end{eqnarray}  

$J_{\rm eff}$ has been calculated in straightforward perturbation theory up to
fifth order \cite{HonL01}, based on the splitting of the general tetramer into
singlet and triplet states we have obtained a result which accounts partly
also for higher orders and allows reasonable estimates for $|J_1-J_3| \ll 1$,
but finite $J_1+J_3$:
\begin{equation}
J_{\rm eff} = \frac{1}{2} \Big\{ \big\{(J_1+J_3-1)^2 + 3 \ (J_1-J_3)^2
   \big\}^{\frac{1}{2}} + J_1 + J_3 
      - \big\{1 + (J_1-J_3)^2\big\}^{\frac{1}{2}} \Big\}.
\end{equation}

\begin{figure}
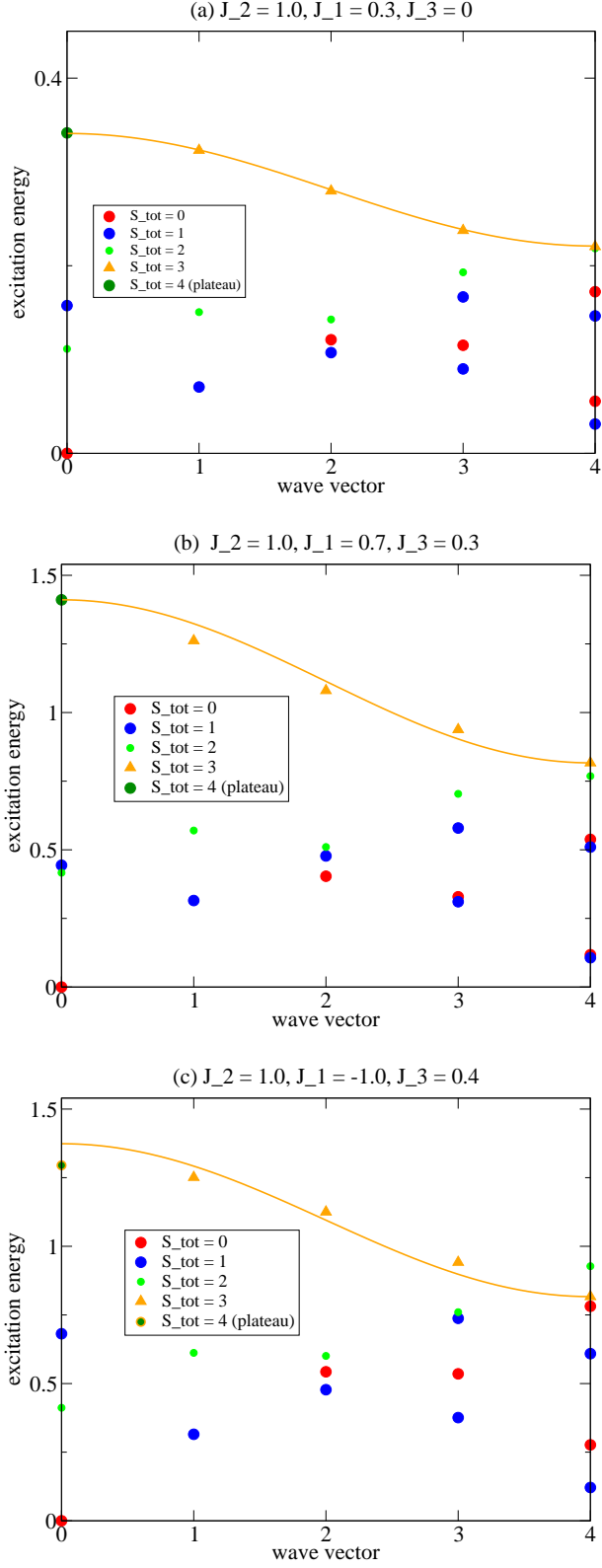

\hspace*{0mm}
\includegraphics[width=80mm]{FiguresMs/low.N24.0300.size.eps}\\[5mm]
\includegraphics[width=80mm]{FiguresMs/low.N24.0703.size.eps}\\[5mm]
\includegraphics[width=80mm]{FiguresMs/low.N24.-1004.size.eps}
\caption{(Color online) Low energy spectrum of the DDC for $N=24, H=0$,
$J_2=1.0$ and (a) $J_1 = 0.3, \ J_3 = 0$, (b) $J_1 = 0.7, \ J_3 = 0.3$, (c)
$J_1 = -1.0, \ J_3 = 0.4$.} 
\label{fig:spectrumlow}
\end{figure}

\begin{figure}
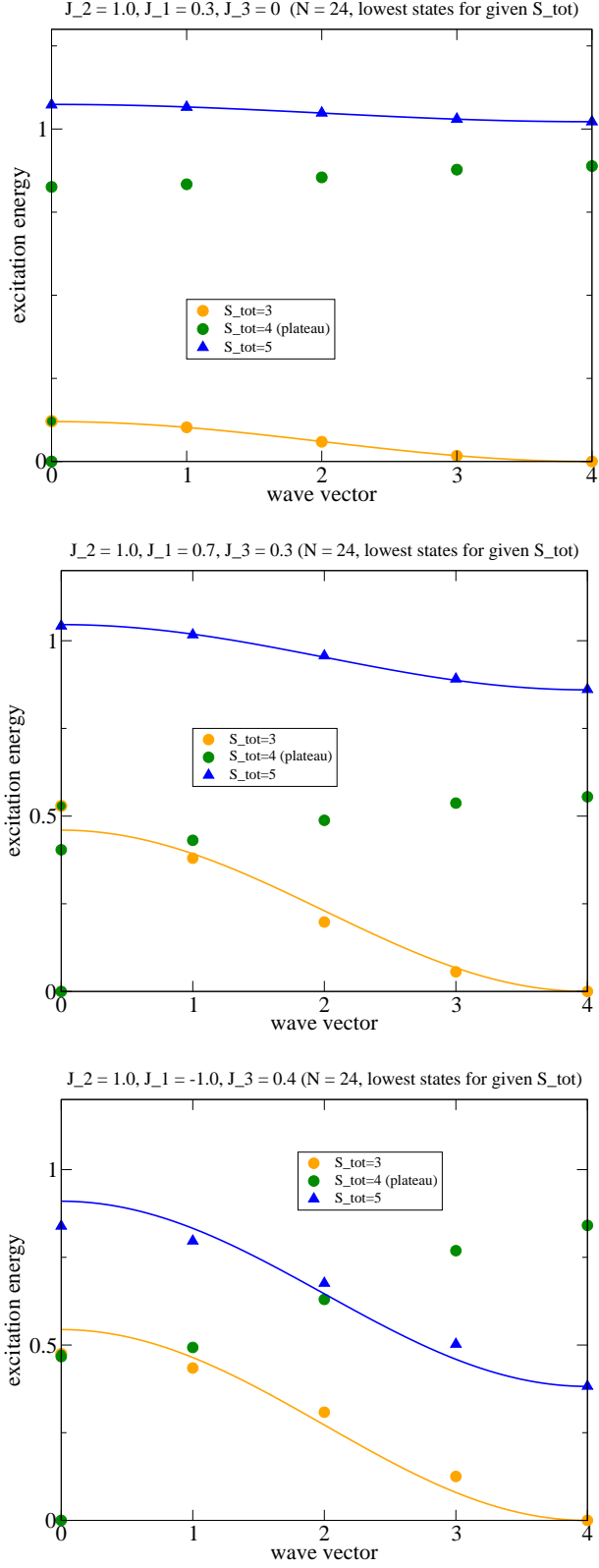

\hspace*{0mm}
\includegraphics[width=80mm]{FiguresMs/field.N24.0300.size.eps}\\[5mm]
\includegraphics[width=80mm]{FiguresMs/field.N24.0703.size.eps}\\[5mm]
\includegraphics[width=80mm]{FiguresMs/field.N24.-1004.size.eps}
\caption{(Color online) Excitation spectrum of the DDC above the 1/3 plateau
for $N=24, H=H_{c1}$, $J_2=1.0$ and (a) $J_1 = 0.3, \ J_3 = 0$, (b) $J_1 =
0.7, \ J_3 = 0.3$, (c) $J_1 = -1.0, \ J_3 = 0.4$.} 
\label{fig:spectrumfield}
\end{figure}

In lowest order perturbation theory the relevant parameter, in addition to the
energy scale set by $J_2$ and to $J_1+J_3$, is the exchange of the effective
HAF determined by $(J_1-J_3)^2$ and many characteristic quantities of the DDC
would be related by simple numerical factors if the mapping were
perfect. Whereas these perturbational results allow to discuss the dynamics in
principle, $J_1, J_3$ values of interest for the bulk of the phase diagram as
well as for a material such as azurite are beyond the validity of perturbation
theory. We therefore present in the following results from the numerical
approaches described above. This will allow us to follow the essential aspects
of the dynamics in the intermediate regime, i.e. through all of the SF
phase. In Figs.~\ref{fig:spectrumlow} and \ref{fig:spectrumfield} we show
excitation spectra for three sets of exchange parameters: set (a) represents
the case of small couplings, set (b) is for a point in the phase diagram close
to the SF to TD transition thought to be qualitatively representative for
azurite \cite{Kik05} and set (c) shows results for the case of one coupling
ferromagnetic. The data are obtained by diagonalizing chains with 24 spins
using the Lanczos algorithm.

The low energy excitation spectra in zero external field are shown in
Fig.~\ref{fig:spectrumlow}. Here and in the following, energies are in units
of $J_2$ and wavevectors in units of $\pi/4$ (for $N=24$, resp. $\pi/3$ for
$N=18$). The spectra include all levels with $S_{\rm tot} = 0,1$ of the spinon
type in the low energy subsystem of 8 spins and in addition the lowest
excitations for $S_{\rm tot} = 2$ (for completeness) and $S_{\rm tot} =
3$. The latter band of excitations is the 'inverted ferromagnon' and a cosine
dispersion approximating the data points is shown for qualitative comparison
to the effective model with its exact cosine dispersion.
Fig.~\ref{fig:spectrumfield} shows the excitation bands in a magnetic field
$H_{c1}$ (beginning of the 1/3 plateau). In magnetic field two Zeeman
components of the $S_{\rm tot} = 4, k=0$ level are relevant: The $S^z_{\rm
tot} = 4$ component turns into the plateau ground state, whereas the $S^z_{\rm
tot} = 3$ component becomes the top of the inverted ferromagnon band, it is
identical to that of Fig.~\ref{fig:spectrumlow} (apart from Zeeman shift) and
now the lowest excitation. In addition, Fig.~\ref{fig:spectrumfield} shows the
lowest excitation band with $S_{\rm tot} = 5$ which requires breaking one
dimer ($J_2$) bond. Cosine dispersions as approximation to the data points are
included for these two bands. For completeness we also show the first full
band with $S_{\rm tot} = 4$ above the plateau ground state.

Among the data shown, the excitations of interest from an experimental point
of view (with large transition matrix elements for e.g. INS) are the spinon
continuum in zero field and the inverted ferromagnon band as well as the
excited dimer band in the plateau field. In addition, in zero field there will
be transitions with energy of the order of $J_2$ to an excited dimer band with
$S_{\rm tot} = 1$ resulting from breaking a $J_2$~bond on top of the effective
chain groundstate. This is more difficult to deal with than the excited dimer
excitation shown in Fig.~\ref{fig:spectrumfield} which is on top of the less
complex saturated effective chain state. We will discuss these excitations
below, based on calculations of all eigenvalues of a $N=18$ chain (see table
III). States with $S_{\rm tot} > 1$ in zero field as well as the band with
$S_{\rm tot} = 4$ above the plateau ground state will be only weakly excited
in INS and analogous experiments: in particular states in the $S_{\rm tot} =
4$ band are obtained from the saturated state by a virtual excitation $\vert s
\uparrow \rangle \to \vert t_+ \downarrow \rangle$.  They have an excited
dimer {\bf and} an overturned spin (compared to the saturated state) in the
low energy subsystem and thus require two spin flips to be excited.

We now discuss how the dynamics changes with varying exchange constants: Set
(a) shows the behaviour typical for the weakly coupled DDC: the bands are well
separated in energy and the cosine dispersion is nearly perfect. Set (b)
displays what is expected for a material such as azurite: in zero field a
spinon continuum should be clearly visible whereas in the plateau regime two
separate bands dominate the picture. The cosine approximation to the
dispersion is less applicable, actually the spectrum of the inverted
ferromagnon is close to linear for smaller wave vectors. For the set (c) which
serves as an example for the alternative suggesting one ferromagnetic coupling
\cite{Hon07}, the dynamics in zero field is seen to be surprisingly close to
that of set (a). This may explain the emergence of the ferromagnetic
alternative from a discussion of static quantities. However, these two sets
lead to strongly differing dynamics in finite field as seen by comparing
Fig.~\ref{fig:spectrumfield} (b) and (c): The standard antiferromagnetc model
(b) implies two well separated bands with rather small widths, whereas the
partly ferromagnetic alternative (c) is characterized by an overlap of the two
bands and a strong dispersion of the excited dimer band.

\begin{figure}
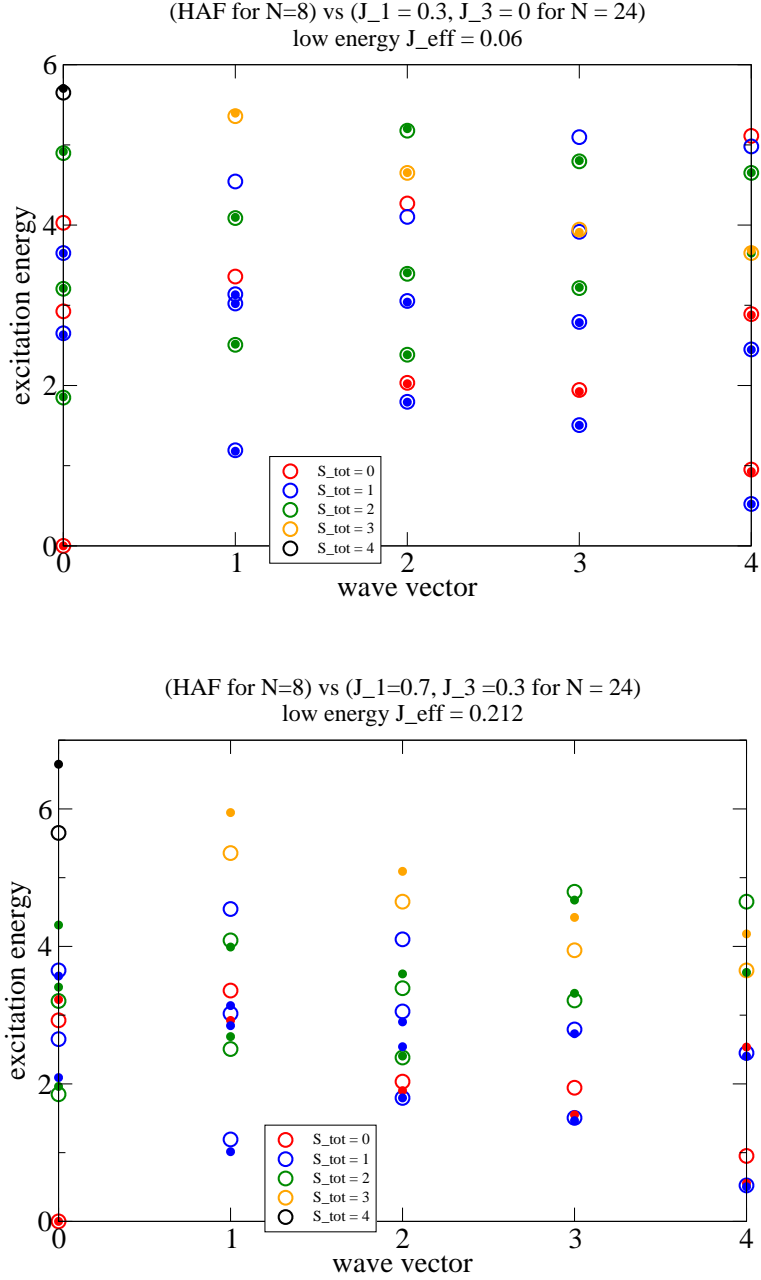

\hspace*{0mm}
\includegraphics[width=100mm]{FiguresMs/haf.N8.0300.size.eps}\\[10mm]
\includegraphics[width=100mm]{FiguresMs/haf.N8.0703.part.size.eps}
\caption{(Color online) Spectrum of the effective HAF (N=8) vs spectrum of the
DDC (N=24) for\\  
(a) $J_1 = 0.3, J_3 = 0$ and (b) $J_1 = 0.7, J_3 = 0.3$.}
\label{fig:compare}
\end{figure}

For a semiquantitative discussion of the low energy dynamics of the DDC the
mapping to the model of an effective HAF is rather useful. Therefore we
discuss shortly the quality of this mapping for $H=0$ in
Fig.~\ref{fig:compare} for the two parameter sets (a) $J_1=0.3, J_3=0$ and (b)
$J_1=0.7, J_3=0.3$. We compare the numerical spectrum for $N=24$ (dots) with
the energies of the $N=8$ HAF with unity exchange constant (open circles). The
energies of the DDC have been scaled by an effective exchange constant $J_{\rm
eff}$, chosen to reproduce the ($N=8$) maximum spinon energy at
$k=\frac{\pi}{2}, S_{\rm tot} = 1$. In (b) energy levels of the $N=8$ HAF
which are beyond the range of the Lanczos calculation for the DDC model have
been omitted from the plot to obtain a clearer picture. Whereas the mapping
for the small parameter values in set (a) is nearly perfect throughout all of
the spectrum, substantial deviations are seen for the parameter set (b): the
low energy spinon part is still reproduced well by the effective model, but
the high energy part, in particular the ferromagnon band with $S_{\rm tot} =
3$ is very different both in energy and in dispersion. A cosine dispersion is
only a rough approximation to the spectrum.

Thus a quantitative experimental investigation of the dynamics will contribute
substantially to locating the position of a specific compound in the phase
diagram of Fig.~\ref{fig:structure+phases}(b). For a quantitative overview
(and possibly use in determination of coupling parameters from experiment) we
reduce in the following the information in these spectra to a few
characteristic numbers to be presented in tables \ref{tab:effective} and
\ref{tab:fields} below. From the numerical data we have calculated values for
the quantities determined by the effective exchange between the quasi-free
spins. We give in table~\ref{tab:effective} numbers for $J_{\rm eff}$
determined from the maximum spinon energy at $k=\pi/2, \ S_{\rm tot} = 1$ (when
multiplied by 1.7964.., the corresponding energy in the N=8 HAF chain, these
numbers lead back to the energy for the DDC model) and for the $S_{\rm tot} =
1$ spinon at $k=\pi$ (gapped due to discreteness). Further we give the width
of the ferromagnon band (which would be $2 J_{\rm eff}$ if the mapping to the
effective model were perfect) and the width of the dimer band above the
plateau with $S_{\rm tot} = N/6 +1$. In table~\ref{tab:fields} we give
numerical values for the characteristic magnetic fields, i.e. beginning
($H_{\rm c1}$) and end ($H_{\rm c2}$) of the plateau as well as the saturation
field $H_{\rm sat}$. In the standard case (actually some exceptions exist
close to the phase transition line) $H_{\rm c1}$ is identical to the
ferromagnon width from table~\ref{tab:effective}. In table~\ref{tab:fields} we
also give the energy scale which is relevant for an application of the
numerical results to azurite: Using the experimental number $H_{\rm sat} = 33
T$ the value of the coupling $J_2$ is calculated from Eq.~(\ref{saturation})
and the values given in table~\ref{tab:fields} (in both $meV$ and $T$) may
serve to obtain energies and fields applying to azurite in standard units. In
tables~\ref{tab:effective} and \ref{tab:fields} three regimes of the SF phase
are covered: (a) values along the $J_3 = 0$ axis (i.e.~for Heisenberg trimer
model) (b) values along a diagonal path which appears as the most interesting
one for discussing azurite and (c) tow examples for ferromagnetic coupling. We
will discuss below the possibility of such an interaction from the point of
view of inelastic excitations. Values along the line $J_1 = 0.6$, passing
through the phase transition at $J_3 \approx 0.364$, will be presented in
section~\ref{sec:phtr-dynamics}.
 
If the mapping to an effective low energy Heisenberg chain were perfect, it
would imply relations between three different quantities which are all
determined by $J_{\rm eff}$:\\ (i) the maximum of the effective one spinon
dispersion at wave vector $\pi/2$: $\epsilon(\pi/2) = \pi/2 J_{\rm eff}$,\\
(ii) the critical field which determines the beginning of the plateau: $H_{\rm
c1} = 2J_{\rm eff}$,\\ (iii) the width of the effective 'inverted ferromagnon'
as the lowest excitation at the beginning of the plateau (in fact, its minimum
defines $H_{\rm c1}$): $\epsilon_{\rm fm}(0) - \epsilon_{\rm fm}(\pi) =
2J_{\rm eff}$. This is automatically equal to (ii) from the definition of
$H_{\rm c1}$, but the cosine dispersion is an additional independent
property. Since the mapping is only approximate, these quantities differ as is
seen in the numerical data and the differences characterize the quality of the
mapping. Actually there are more possibilities to extract $\rm J_{eff}$ from
the numerical data such as the energy of the lowest spinon singlet at $k=\pi$
and the ground state energy (suitably extracted from the energy of the
saturated subsystem state), but numbers from these approaches essentially
confirm the picture as it has emerged from the tables above. The essential
conclusion for the real DDC is that the effective coupling $\rm J_{eff}$
depends on energy.

\begin{table}
\begin{tabular}{| c | c |c|c|c|} \hline
\quad couplings $J_1, J_3$ \quad & $J_{\rm eff}$ from & $J_{\rm eff}$ from & 
       \quad ferromagnon \quad & \quad dimer band \quad \\[-2mm] 
   & \quad spinon maximum \quad  & \quad spinon at $k=\pi$ \quad 
                           & width & width \\ \hline \hline
0.02, \quad 0.00  & 2.06 10$^{-3}$ & 2.06 10$^{-3}$ 
                     & 4.12 10$^{-3}$ & 2.02 10$^{-3}$ \\ \hline 
0.30, \quad 0.00  & 0.060 & 0.060 & 0.121 & 0.051 \\ \hline 
0.60, \quad 0.00  & 0.223 & 0.207 & 0.480 & 0.218 \\ \hline 
0.80, \quad 0.00  & 0.343 & 0.291 & 0.795 & 0.400 \\ \hline \hline
0.70, \quad 0.30  & 0.212 & 0.204 & 0.529 & 0.181 \\ \hline 
0.65, \quad 0.25  & 0.192 & 0.186 & 0.452 & 0.152 \\ \hline 
0.60, \quad 0.20  & 0.172 & 0.168 & 0.388 & 0.132 \\ \hline 
0.40, \quad 0.00  & 0.107 & 0.106 & 0.220 & 0.080 \\ \hline \hline
0.40, \quad -1.00 & 0.266 & 0.232 & 0.478 & 0.456 \\ \hline 
0.02, \quad -0.40 & 0.046 & 0.046 & 0.091 & 0.068 \\ \hline
\end{tabular}
\caption{DDC parameters related to the effective interaction (see text)}
\label{tab:effective}
\end{table}

\begin{table}
\begin{tabular}{| c | c |c|c|c|} \hline
\quad couplings $J_1, J_3$ \quad & \phantom{hhh} $H_{c1}$ \phantom{hhh} 
       & \phantom{hhh} $H_{c2}$ \phantom{hhh} 
       & \phantom{hhh} $H_{\rm sat}$ \phantom{hhh} 
       & \phantom{hhh} energy scale $J_2$ in \phantom{hhh} \\[-2mm] 
   &  &  &  & $meV$ resp. $T$ \\ \hline \hline
0.02, \quad 0.00  & 4.12 10$^{-3}$ & 1.0100 & 1.0102 
                      & $\quad 4.16 \ meV \equiv 32.7 \ T \quad$ \\ \hline 
0.30, \quad 0.00  & 0.121 & 1.143 & 1.200 
                      & $\quad 3.50 \ meV \equiv 27.5 \ T \quad$ \\ \hline 
0.60, \quad 0.00  & 0.480 & 1.245 & 1.500 
                      & $\quad 2.80 \ meV \equiv 22.0 \ T \quad$ \\ \hline 
0.80, \quad 0.00  & 0.795 & 1.272 & 1.740 
                  & $\quad 2.41 \ meV \equiv 19.0 \ T \quad$ \\ \hline \hline
0.70, \quad 0.30  & 0.529 & 1.390 & 1.627 
                      & $\quad 2.58 \ meV \equiv 20.3 \ T \quad$ \\ \hline 
0.65, \quad 0.25  & 0.452 & 1.371 & 1.569 
                      & $\quad 2.68 \ meV \equiv 21.0 \ T \quad$ \\ \hline 
0.60, \quad 0.20  & 0.388 & 1.342 & 1.512 
                      & $\quad 2.78 \ meV \equiv 21.8 \ T \quad$ \\ \hline 
0.40, \quad 0.00  & 0.220 & 1.183 & 1.29    
              & $\quad 3.26 \ meV \equiv 25.6 \ T \quad$ \\ \hline \hline 
0.40, \quad -1.00 & 0.478 & 0.861 & 1.234 
                      & $\quad 3.40 \ meV \equiv 26.7 \ T \quad$ \\ \hline 
0.02, \quad -0.40 & 0.091 & 0.824 & 0.868 
                      & $\quad 4.84 \ meV \equiv 38.0 \ T \quad$ \\ \hline 
\end{tabular}
\caption{DDC parameters related to the characteristic fields and unit of
                      energy for azurite (see text)}
\label{tab:fields}
\end{table}

\begin{figure}
\hspace*{0mm}
\includegraphics[width=100mm]{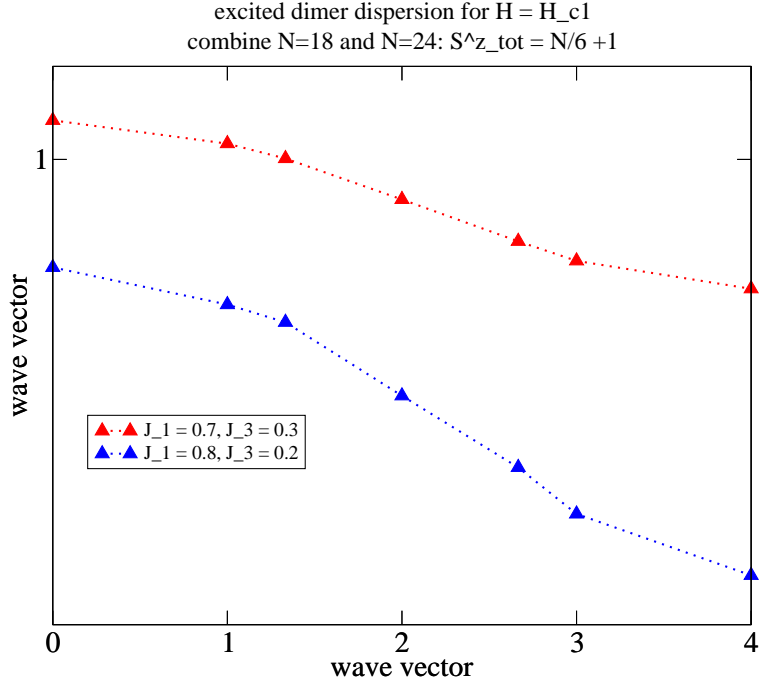}
\caption{(Color online) Excited dimer band of the DDC (N=24 and N=18) above
the 1/3 plateau ground state (in a magnetic field $H_{c1}$) for $J_1 = 0.7,
J_3 = 0.3$ and $J_1 = 0.8, J_3 = 0.2$} 
\label{fig:07.08.plateau}
\end{figure}

The quantities in tables \ref{tab:effective} and \ref{tab:fields} have been
calculated for $N=24$, but a comparison with results for $N=12$ and $N=18$
shows that finite size effects are very small, indeed negligible to the
accuracy given. This is due to the fact that e.g.~$J_{\rm eff}$ is determined
by comparing two finite systems, the HAF chain with $N=8$ and the DDC with
$N=24$. This is illustrated in Fig.~\ref{fig:07.08.plateau} where the excited
dimer bands for two different sets of couplings are shown combining results
for $N=24$ and $N=18$ in the same graph. We conclude that the the effective
parameters considered so far can be reliably determined for the infinite
chain. The situation is different for the last quantity of interest, the
dynamics of the one dimer excitations for $H=0$. These excitations originate
by exciting one dimer from singlet to triplet on top of the spinon continuum
of the effective HAF chain (instead of on top of the saturated effective HAF
chain for $\Delta_{\rm dimer}$ as given above). In the following we give
results on this zero field branch as far as numerically possible:

At zero field, the excited dimer states form a continuum as well and the
infinite chain limit can only be obtained by extrapolation in $1/N$. With
$N=12$ and $N=18$ as the only available numbers of spins the extrapolation can
only be done for wavevectors $k=0$ and $k=\pi$. The lowest energy levels in
our finite systems result from coupling of spinons at wavevectors 0 (singlet
only) and $\pi$ (singlet and triplet) to the dimer triplet, i.e. we get a band
of one singlet, three triplets and one quintuplet. However, the separation
both between these multiplets and to the higher levels is due to the finite
size, whereas in the thermodynamic limit a continuum with energy of the order
of $J_2$ will result. This is to some extent reflected in the results of
finite size extrapolation of the multiplet energies related to one excited
dimer: There is a clear tendency for the energies to converge to the same
value. Thus only one energy in this high energy subspace can be given
reliably, no reliable information can be obtained in this approach about
splitting into bands. In table~\ref{tab:exciteddimer} we give the energy of
the lowest state in the continuum of excited dimers obtained this way for a
number of coupling constants.

\begin{table}
\begin{tabular}{| c | c | c |} \hline
\quad couplings $J_1, J_3$ \quad 
       & \phantom{hhh} $\epsilon^{\rm dimer}(k=0)$ \phantom{hhh} 
       & \phantom{hhh} $\epsilon^{\rm dimer}(k=\pi)$  \phantom{hhh}\\[-5mm] 
       &  & \\ \hline \hline
0.30, \quad 0.00  & 0.861 & 0.893 \\ \hline 
0.60, \quad 0.05  & 0.988 & 0.978 \\ \hline 
0.60, \quad 0.25  & 0.720 & 0.711 \\ \hline 
0.65, \quad 0.25  & 0.854 & 0.793 \\ \hline 
0.70, \quad 0.30  & 0.864 & 0.776 \\ \hline 
0.40, \quad -1.00 & 1.226 & 1.255 \\ \hline 
\end{tabular}
\caption{Energies $\epsilon(k=0)$ and $\epsilon(k=\pi)$ in the DDC lowest 
excited dimer band (onset of the continuum). Values extrapolated from $N=12$
and $N=18$ to infinite $N$}
\label{tab:exciteddimer}
\end{table}

\section{Dynamics in the dimer-tetramer phase}
\label{sec:td-dynamics}

The ground state in the tetramer-dimer phase is twofold degenerate and
develops from the ground state on the symmetry line $J_1=J_3$. On this line
the two ground states can be written down explicitly (even for finite systems
with an arbitrary even number of cells of 3 spins). They are given by the
alternating sequence of the dimer singlet $\mathbf S$ and the lowest
tetramersinglet $\mathbf T$. This allows the two equivalent configurations
\begin{equation}  
{\bf \dots S \quad T \quad  S \quad T \quad  S \quad T \dots} \qquad
{\rm and}  \qquad
{\bf \dots T \quad  S \quad T \quad  S \quad T \quad  S \dots}
\label{eq:tdgroundstates}
\end{equation}  
describing the two degenerate ground states.

The lowest excitations above these ground states are obtained as solitons which
are defined by gluing together the two degenerate ground states in a localized
region on the chain. This gives e.g.~the state
\begin{equation}  
{\bf \dots S \quad T \quad  S \ * \  S \quad T \dots}
\label{eq:soliton}
\end{equation}  
where ${\bf *}$ denotes a free spin. A soliton is possible only with two dimer
singlets adjacent to each other and a free spin between them whereas a
configuration ${\bf \dots T \quad T \dots}$ obviously does not exist. On the
symmetry line $J_1=J_3$ there are $N/3$ degenerate localized one-soliton
configurations. They start to propagate and to form a soliton band for $J_1
\ne J_3$. The properties of a single soliton can suitably be investigated for
chains with an odd number of cells $N/3$ when periodic boundary conditions
require the existence of one soliton in the ground state. The hopping
process 
\begin{equation}  
{\bf  T \quad  S \ * \  \quad  \to \quad \ * \ S \quad T }
\label{eq:hopping}
\end{equation}  
with amplitude $t$ leads to the propagation of solitons and the formation
of a ground state band with energy
\begin{equation}  
\epsilon^{\rm sol}(k) = E_0 \ + \ 2t \cos 2k.
\label{eq:solitondisp}
\end{equation}  

Thus, from numerical calculations for an odd number of cells the hopping
amplitude is easily determined even when only small systems are available. For
infinite chains with periodic boundary conditions and an even number of cells
$N/3$ which possess two degenerate ground states and enforce an even number of
solitons, the low energy spectrum is dominated by the two soliton continuum
emerging from independent propagation of two solitons with wavevectors
$\frac{1}{2}k + k_1$ and $\frac{1}{2}k - k_1$, i.e.
\begin{equation}  
\epsilon^{\rm 2sol}(k) = 2E_0 \ + \ 4t \ \cos k \ \cos 2k_1.
\label{eq:twosolitondisp}
\end{equation}  

\begin{figure}
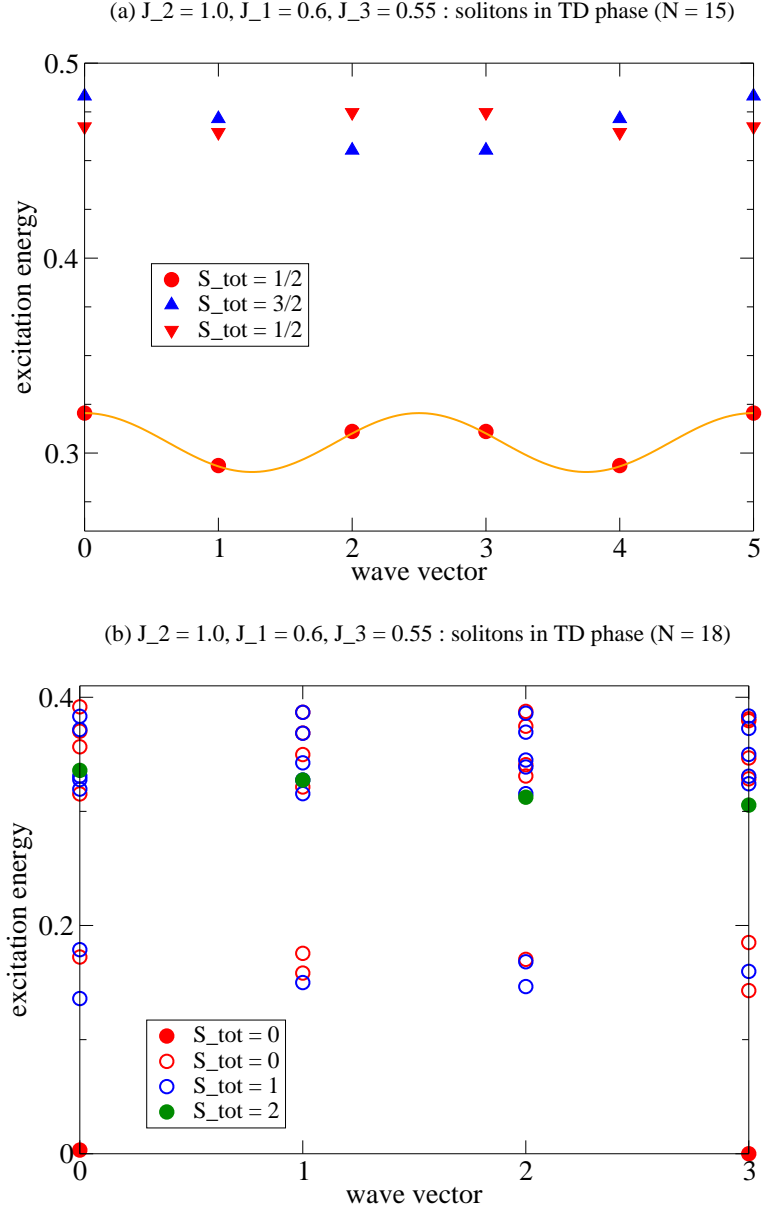

\hspace*{0mm}
\includegraphics[width=100mm]{FiguresMs/solitons.N15.60.55.size.eps}\\[5mm]
\includegraphics[width=100mm]{FiguresMs/solitons.N18.60.55.size.eps}
\caption{(Color online) Soliton spectrum of the DDC ((a) N=15, (b) N=18) in
the TD phase for $J_1 = 0.6, J_3 = 0.55$. The N=15 spectrum is shown in the
complete Brillouin zone $0 \dots 2\pi$ to demonstrate the $\cos 2k$ dispersion
of the single soliton.}
\label{fig:tdsolitons}
\end{figure} 

Numerically obtained soliton spectra are shown in Fig.~\ref{fig:tdsolitons}
for (a) $N=15$ and (b) $N=18$ for the point $J_1 = 0.6, J_3 = 0.55$ close to
the symmetry line in the phase diagram. $\epsilon^{\rm sol}(k)$ in
Fig.~\ref{fig:tdsolitons}(a) clearly shows the $\cos 2k$ dispersion, whereas
the dispersion of $\epsilon^{\rm 2sol}(k)$ in Fig.~\ref{fig:tdsolitons}(b) is
somewhat more complicated due to the small size of the system.  Also shown are
the three, resp. four-soliton bands demonstrating the clear division of the
spectra into distinct solitons bands for this nearly symmetric set of
couplings (the zero of energy in Fig.~\ref{fig:tdsolitons}(a) is taken from
the noninteracting limit). The situation is analogous to the Ising chain with
small transverse interactions, the system where the dynamics of magnetic
solitons was discussed first \cite{Vil75,IshS80}. Slightly different, the
soliton spin 1/2 here is a real spin 1/2 which can be attributed to the free
electron of the Cu$^{2+}$~ion between the two dimers forming the domain
wall. For small numbers $N/3$ the soliton spectrum clearly shows the effects
of the different symmetry of singlet and triplet.

From the one soliton data for $N=15$ and $J_1 = 0.6, J_3 = 0.55$ the hopping
amplitude is deduced as $t \approx 0.0076$. For the two soliton data for
$N=18$ the corresponding calculation has to include the possibility of two
neighboring solitons as well as the resulting symmetry effects and gives a
somewhat higher value, $t \approx 0.0097$. The deformation of the zero order
wave function due to neighboring solitons is strongest for small systems which
explains the difference. However, for $N=24$ (when only the lowest 2 soliton
band is accessible in Lanczos calculations) we obtain $t \approx 0.0076 \pm
0.0002$, identical to the one soliton result for $N=15$ within the uncertainty
resulting from matching the cosine dispersion for the different wave vectors.

\section{Crossing the phase transition line}
\label{sec:phtr-dynamics}

In this section we present some data for the specific heat and for the
spectrum of low-lying excitations in order to approach the behavior of the
system when its coupling constants change between well defined end points, one
in the SF phase, the other one in the TD phase, thus crossing the phase
transition line. Evidently, owing to the small system sizes accessible only in
our calculations, we cannot claim that these data describe correctly the most
interesting aspect, namely the critical behavior; on the other hand our data
for both the specific heat and the spectrum of low-lying excitations set a
reasonable frame for the transition regime, to be filled by more detailed
calculations later.  In the following we present results for the DDC on the
line $J_2 = 0.6$ for varying $J_3$. As discussed above, the phase transition
along this line is of Kosterlitz Thouless type and can be considered as a
generalization of the phase transition in the HAF with both nearest and next
nearest neighbor exchnage. We therefore have applied the procedure of
Ref.~\onlinecite{OkaN92} to determine the critical coupling and find that the
phase transition occurs at $J_3 \approx 0.364$.

\begin{figure}
\hspace*{0mm}
\includegraphics[width=80mm]{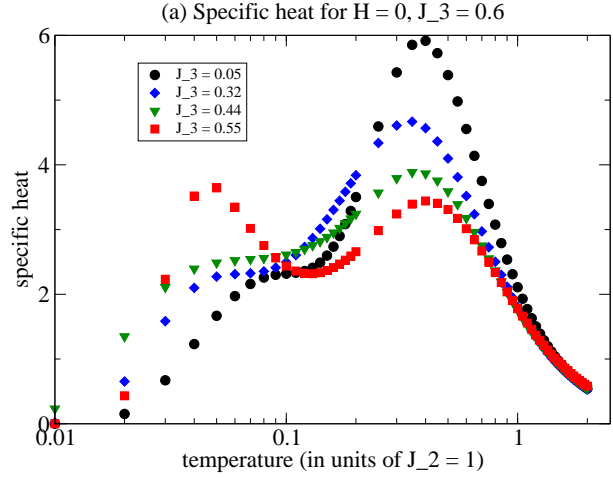}\\[12mm]
\includegraphics[width=80mm]{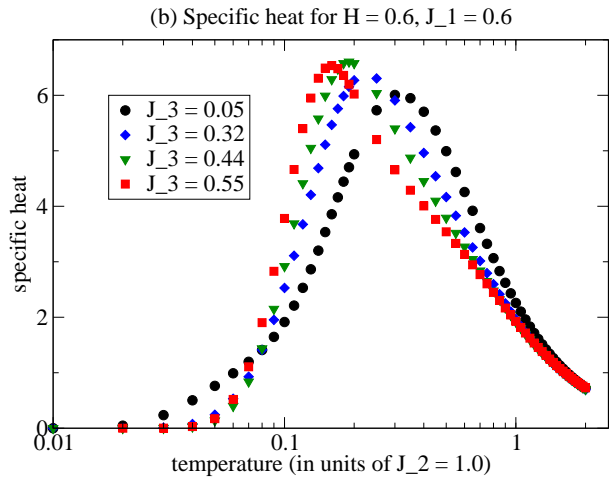}\\[12mm]
\includegraphics[width=80mm]{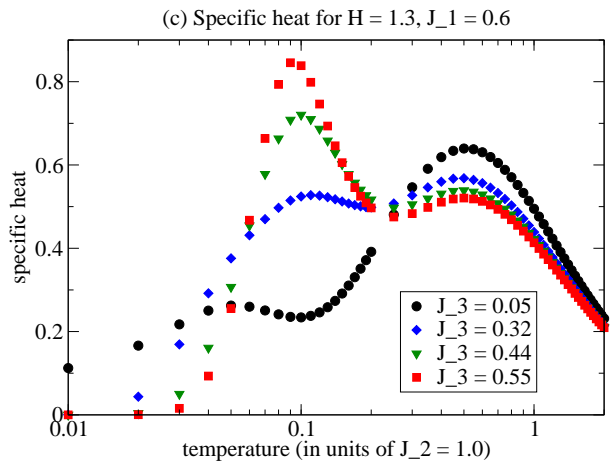}\\[1mm]
\caption{(Color online) Specific heat of the DDC (from all levels for $N=18$)
for $H=0$ (a), $H=0.6$ (b) and $H=1.3$ (c), fixed $J_1 = 0.6$ and varying
$J_3$ through the phase transition.}
\label{fig:pt:specheat}
\end{figure}

Fig.~\ref{fig:pt:specheat} shows a sequences of specific heat data varying
$J_3$ for fixed $J_1$ for three different values of the magnetic field.  The
data are obtained from the full spectrum for the $N=18$ chain and therefore
cover reliably the complete temperature regime although critical properties
near the critical coupling $J_3 \approx 0.364$ will appear smeared out. In
all diagrams we use a logarithmic temperature scale adequate to the strongly
different energy scales. For all magnetic fields the specific heat exhibits a
high temperature peak at $T \approx 0.5$, whereas the low temperature
properties reflect the structure of the system: For $H=0$ and low temperature
(Fig.~\ref{fig:pt:specheat}a) the SF phase is characterized by a continuously
increasing contribution to the specific heat, the remnant of the effective
Luttinger liquid. With increasing $J_3$ this contribution develops gradually
into the gapped contribution of the TD phase with characteristic shoulders on
both sides of the phase transition at $J_3 \approx 0.364$. For $H=0.6$
(Fig.~\ref{fig:pt:specheat}b) the DDC always is in the gapped plateau
regime and the specific heat shows little variation with the coupling. For
$H=1.3$ (Fig.~\ref{fig:pt:specheat}c) the DDC is close to saturation and
the differences between the two phases, resulting from the energy spectra
above the plateau gap, become apparent again. In particular, the well defined
grouping into soliton bands at $J_3 = 0.55$ leads to a strong low temperature
peak which actulaly develops continuously from lower $J_3$ values. It would be
interesting to see, using more powerful numerical methods, whether the
development of this peak in the nearly saturated case shows critical
properties.

\begin{figure}
\hspace*{0mm}
\includegraphics[width=120mm]{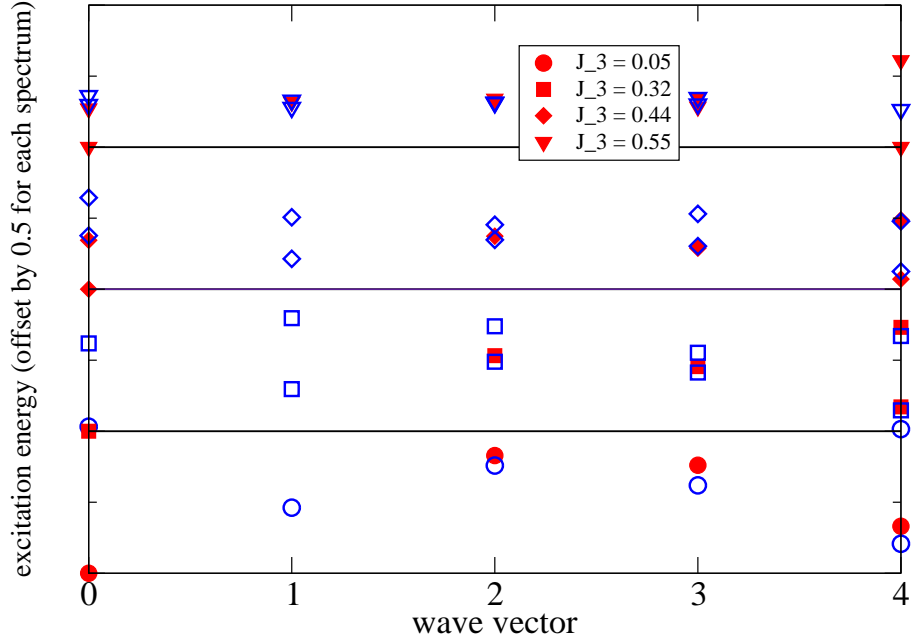}\\[5mm] 
\caption{(Color online) Low-lying spectra ($S_{\rm tot} = 0$ (red/full) and
$S_{\rm tot} = 1$ (blue/open)) of the DDC (N=24) for $J_1 = 0.6$ and (from top
to bottom) $J_3 = 0.55, 0.44, 0.32, 0.05$. The energy range for each spectrum
is 0.5.}  
\label{fig:pt:spectra}
\end{figure} 

Fig.~\ref{fig:pt:spectra} shows a sequence of spectra of low-lying excitations
($S_{\rm tot} = 0, 1$) in zero magnetic field obtained by the same procedure
as the spectra shown in sections \ref{sec:sf-dynamics} and
\ref{sec:td-dynamics}. Qualitatively, the transition from the SF
phase with its gapless spinon continuum to the gapped soliton spectra in the
TD phase is clearly seen qualitatively, as far as possible for the limited
size of the system: with increasing $J_3$ the second degenerate ground state
at $k = \pi$ emerges, the spinon continuum is compressed
into the soliton band and the gapless character disappears. Excitation
spectra in finite magnetic field, in particular for fields in the plateau
regime, on the other hand do not show specific variations but rather
continuous changes across the phase diagram without prominent features close
to $J_3 = 0.364$, the location of the phase transition in zero field. This
illustrates, as expected, that the signature of the phase transition in the
dynamics is limited to zero external field and low-lying excitations wehreas
the 1/3 plateau as well as the accompanying excitations are continuous across
the phase diagram.

\section{Conclusions}
\label{sec:conclusion}

For the $S=1/2$ distorted diamond chain in both the spin fluid (SF) and the
tetramer-dimer (TD) phases we have calculated the spectra of low-lying
excitations and the specific heat. We have used both full numerical
diagonalization (for chains with up to 18 spins) and the Lanczos algorithm
(for chains with up to 24 spins) and have discussed the results in relation to
approximate analytic approaches. Except close to the SF-TD phase transition
results for our small systems are shown to represent the thermodynamic
limit. Our calculations are for arbitrary value of the external magnetic
field, results are mainly given for zero field and for fields corresponding to
the 1/3 plateau regime.

In the SF phase the low energy spectra can be related to a Heisenberg
antiferromagnetic chain with effective interaction $J_{\rm eff}$. For
parameters beyond the validity of a perturbative approach, this effective
interaction has to be allowed to be energy dependent. The lowest excitations
in the plateau regime are the inverted ferromagnon and the propagating single
dimer triplet excitation with, however, partly strong modifications of the
corresponding cosine dispersions. The values of the characteristic parameters
($J_{\rm eff}$, extent of the plateau regime, widths of the cosine bands) are
given for typical paths crossing the SF phase. These data should allow to
decide whether a material such as Cu$_3$(CO$_3$)$_2$(OH)$_2$ (azurite) is
sufficiently well described by the DDC model and, if so, to determine the
corresponding couplings. The standard assumption for azurite is to take all
couplings as antiferromagnetic and we have shown that the spectra of low-lying
excitations exhibit large and characteristic changes when the possibility of
one ferromagnetic coupling is introduced. We therefore expect that our data
will allow to interpret quantitatively experimentla data on azurite. This
refers in particular to the results of inelastic neutron scattering
experiments \cite{RulST07}. Considering the present status of such
investigations, our results do not confirm the conclusion of at least one
ferromagnetic coupling in azurite. Generally, our results lead us to describe
the following signatures when ferromagnetic couplings are present:\\ 
(i) Whereas the dimer width is roughly 1/2 of the ferromagnon width for af
couplings (as suggested by perturbation theory), for ferromagnetic couplings
these widths tend to become equal.\\
(ii) The sign of the couplings has a marked influence on the relative
appearance of the ferromagnon and the excited dimer band: For couplings $J_1,
J_3 = (-1, 0.4)$ these bands above the 1/3 plateau overlap, whereas
for $(-0.4, 0.02)$ ferromagnon a swell as dimer width become very small 
and correspondingly $H_{c1}$ becomes much smaller than in perturbation theory.

The low-lying excitations in the TD phase with its twofold degenerate ground
state are shown to be solitons. The width of the one soliton band as
determined from the $N=15$ chain not too far from the symmetry line $J_1 =
J_3$ reproduces well the soliton bands in the $N=24$ chain and therefore gives
reliably the tunneling amplitude for the soliton propagation in the
thermodynamic limit.

We have also shown spectra as well as the specific heat on a line across the
SF-TD phase transition. Although the small systems accessible to us do not
allow to discuss critical properties of the DDC close to this
Kosterlitz-Thouless transition, the variation of the dynamical properties
through the transition become clear. In particular, only the low energy
properties, determining the behavior of the system at zero field, carry the
signature of the phase transition.

Generally, the DDC has many features in common with the antiferromagnetic
Heisenberg chain with nearest and next-nearest exchange and the SF-TD phase
transition is of the same type as the KT transition at $J_{\rm NNN} =
0.2411... J_{\rm NN}$ in this system. On the other hand, we have shown that
the additional degrees of freedom, resulting from the possibility to excite
the $J_2-$ dimers to the triplet state, show up clearly in the dynamics. We
leave to the future to investigate the influence of these degrees of frededom
on the phase transition using more powerful anlytical and numerical methods.

\section*{acknowledgements} 

We wish to thank H.~Ohta, H.~Kikuchi, K.~Rule, S.~S\"ullow and
D.~A.~Tennant for stimulating discussions. We gratefully acknowledge that
computational facilities for the numerical calculations were generously
provided by the John von Neumann-Institut for Computing at J\"ulich Research
Center.


\begin{thebibliography}{50}
\section*{References}

\bibitem{TakKS96} K.~Takano, K.~Kubo and H.~Sakamoto, J.~Phys.~Condens.~Matter
{\bf 8}, 6405 (1996)
  
\bibitem{OkaT99} K.~Okamoto T.~Tonegawa, Y.~Takahashi and M.~Kaburagi,
J.~Phys.~Condens.~Matter {\bf 11}, 10485 (1999)

\bibitem{TonO2000} T.~Tonegawa, K.~Okamoto, T.~Hikihara, Y.~Takahashi and
M.~Kaburagi, J.~Phys.~Soc.~Jpn {\bf 69} Suppl.A, 332 (2000)

\bibitem{Kik05} H.~Kikuchi, Y.~Fujii, M.~Chiba, S.~Mitsudo, T.~Idehara,
T.~Tonegawa, K.~Okamoto, T.~Sakai, T.~Kuwai and H.~Ohta, Phys.~Rev.~Letters
{\bf 94}, 227201 (2005)  

\bibitem{Oht03} H.~Ohta, S.~Okubo, T.~Kamikawa, T.~Kunimoto, Y.~Inagaki,
H.~Kikuchi, T.~Saito, M.~Azuma and M.~Takano, J.~Phys.~Soc.~Jpn {\bf 72}, 2464
(2003) 

\bibitem{HonL01} A.~Honecker and A.~L\"auchli, Phys.~Rev. B{\bf 63}, 174407
  (2001)

\bibitem{OkaT03} K.~Okamoto, T.~Tonegawa and M.~Kaburagi,
J.~Phys.~Condens.~Matter {\bf 15}, 5979 (2003)

\bibitem{Hon07} B.~Gu and B.~Su, Phys.~Rev.~Lett. {\bf 97}, 089701 (2006), 
                A.~Honecker, private communication (2007)

\bibitem{RulST07} K.C.~Rule, A.U.B.~Wolter, S.~S\"ullow, D.A.~Tennant,
A.~Br\"uhl, S.~K\"ohler, B.~Wolf, M.~Lang and J.~Schreuer, preprint cond-mat
arXiv:0709.2560 (2007) 

\bibitem{Vil75} J.~Villain, Physica B {\bf 79}, 1 (1975) 

\bibitem{IshS80} N.~Ishimura and H.~Shiba, Progr.~theor.~Phys.~(Osaka) {\bf
63}, 743 (1980) 

\bibitem{OkaN92} K.~Okamoto and K.~Nomura, Phys.~Lett. A {\bf 169}, 433 (1992)
\end{thebibliography}
\end{document}